**Original Article:** Assessing Selection Bias in Regression Coefficients Estimated from Non-Probability Samples, with Applications to Genetics and Demographic Surveys

**Brady T. West (Corresponding Author)[1]**
Survey Research Center, Institute for Social Research
University of Michigan-Ann Arbor
426 Thompson Street, Ann Arbor, MI, USA, 48106-1248
Phone: 734-647-4615, Fax: 734-647-2440
Email: bwest@umich.edu

**Roderick J.A. Little**
Department of Biostatistics, School of Public Health
University of Michigan-Ann Arbor
Email: rlittle@umich.edu

**Rebecca R. Andridge**
Division of Biostatistics, College of Public Health
Ohio State University
Email: andridge.1@osu.edu

**Philip S. Boonstra**
Department of Biostatistics, School of Public Health
University of Michigan-Ann Arbor
Email: philb@umich.edu

**Erin B. Ware**
Survey Research Center, Institute for Social Research
University of Michigan-Ann Arbor
Email: ebakshis@umich.edu

**Anita Pandit**
Department of Biostatistics, School of Public Health
University of Michigan-Ann Arbor
Email: anitapan@umich.edu

**Fernanda Alvarado-Leiton**
Michigan Program in Survey Methodology, Institute for Social Research
University of Michigan-Ann Arbor
Email: mleiton@umich.edu

[1] This work was supported by grants from the National Institutes for Health (#1R21HD090366-01A1, #R01AG055406). The National Survey of Family Growth (NSFG) is conducted by the Centers for Disease Control and Prevention's (CDC's) National Center for Health Statistics (NCHS), under contract # 200-2010-33976 with University of Michigan's Institute for Social Research with funding from several agencies of the U.S. Department of Health and Human Services, including CDC/NCHS, the National Institute of Child Health and Human Development (NICHD), the Office of Population Affairs (OPA), and others listed on the NSFG webpage (see http://www.cdc.gov/nchs/nsfg/). The views expressed here do not represent those of NCHS nor the other funding agencies.

**Abstract**

Selection bias is a serious potential problem for inference about relationships of scientific interest based on samples without well-defined probability sampling mechanisms. Motivated by the potential for selection bias in (a) estimated relationships of polygenic scores (PGSs) with phenotypes in genetic studies of volunteers, and (b) estimated differences in subgroup means in surveys of smartphone users, we derive novel measures of selection bias for estimates of the coefficients in linear and probit regression models fitted to non-probability samples, when aggregate-level auxiliary data are available for the selected sample and the target population. The measures arise from normal pattern-mixture models that allow analysts to examine the sensitivity of their inferences to assumptions about non-ignorable selection in these samples. We examine the effectiveness of the proposed measures in a simulation study, and then use them to quantify the selection bias in (a) estimated PGS-phenotype relationships in a large study of volunteers recruited via Facebook, and (b) estimated subgroup differences in mean past-year employment duration in a non-probability sample of low-educated smartphone users. We evaluate the performance of the measures in these applications using benchmark estimates from large probability samples.

**Key Words:** Linear Regression, Probit Regression, Non-Probability Samples, Selection Bias, Polygenic Scores, National Survey of Family Growth

# 1. Introduction

The random selection of elements from a finite population of interest into a probability sample, where all population elements have a known non-zero probability of selection, ensures that elements included in the sample, appropriately weighted if necessary, mirror the target population in expectation. That is, for all variables of interest, the mechanism of selection of a subset of elements into the sample is ignorable, following the theoretical framework for missing data mechanisms originally introduced by Rubin (1976).

Unfortunately, the modern survey research environment presents substantial challenges for probability sampling: sampled units are harder to contact, survey response rates continue to decline (Brick and Williams, 2013; de Leeuw, Hox and Luiten., 2018; Williams and Brick, 2018), and the costs of collecting and maintaining scientific probability samples are steadily rising (Presser and McCulloch, 2011). Given these challenges, researchers are turning to the collection and analysis of data from non-probability samples. These data may be scraped from social media platforms, collected from commercial databases, gathered from online searches, or recorded via online surveys of volunteers (Baker et al., 2013). In clinical trials, inferences about the effects of treatments in a target population are nearly always based on volunteer samples. The protection of ignorable selection conveyed by probability sampling no longer applies in these settings, and probability samples with very low response rates in probability samples raise similar concerns. Classical design-based methods of survey inference about finite target populations do not apply, and model-based inferential methods for non-probability samples are an active focus of current survey research (Elliott and Valliant, 2017; Valliant, 2019).

There is thus a critical need for diagnostic measures to both assess and correct for the bias in estimates from non-probability samples. Nearly all the work in this area has focused on measuring the potential bias in estimates of means and proportions. Nishimura, Wagner and Elliott (2016) demonstrated that existing measures did not do a good job in detecting the selection bias in descriptive estimates introduced by non-ignorable survey nonresponse. Little et al. (2020) and Andridge et al. (2019) proposed new measures of bias to address this deficiency, based on adjustments for nonignorable nonresponse in Andridge and Little (2011). These measures outperformed alternative diagnostic measures such as the R-indicator (Schouten, Cobben and Bethlehem, 2009) in simulation studies (Boonstra et al., 2021).

Selection bias can also affect estimates of the relationships between variables. In particular, good measures are needed of the extent to which estimates of regression coefficients from a non-probability sample are subject to bias due to non-ignorable selection. We consider this question in the context of two motivating examples, where benchmark data are available to measure the actual degree of selection bias in the regression coefficients estimated from the non-probability sample.

The first setting concerns relationships between the polygenic score (PGS; Ware et al., 2017) -- a summary of several thousand genetic measures available for a given individual -- and selected phenotypes. These relationships are often estimated based on large samples of volunteers and hence are subject to potential non-ignorable selection bias. In the second setting, survey researchers are often interested in subgroup differences in estimated descriptive parameters (e.g., employment rates) and turning to data collection using smartphones, given the rapidly increasing prevalence of these mobile devices (see https://tinyurl.com/yaeg3rwn). However, smartphone users are not a random sample of the population (Couper et al., 2018), introducing concerns about selection bias in these estimated subgroup differences. More specifics of these two applications are given in Section 2.

In Section 3, we extend the measures for estimates of means and proportions in Little et al. (2020) and Andridge et al. (2019) to the coefficients in linear and probit regression models. In Section 4, we assess the ability of these measures to detect selection bias in a simulation study, and we then apply them to our two motivating applications in Section 5. Section 6 presents conclusions and topics for future research.

## 2. Motivating Applications

*2.1 Polygenic Score-Phenotype Relationships in the Genes for Good Study*

Genes for Good (GfG) is a research study based at the University of Michigan that seeks to engage the public in genetic research. Volunteers 18 years of age and above currently living in the United States enroll in the study via a Facebook app, which serves as a tool for them to engage in all aspects of the study (including the answering of health-related survey questions). Volunteers consent to be genotyped and provide saliva samples via mail after answering a minimum number of surveys. Researchers use the resulting genetic profiles to investigate the effects of certain genetic variants on health measures that volunteers self-report via the app. The study is based entirely on volunteers, of which there have been more than 77,000 to date (20,100 of which had been genotyped at the time of this analysis), and therefore does not have an underlying probability sampling mechanism. One can find additional details on the GfG study at https://genesforgood.org.

Polygenic Scores (PGSs), or genetic risk scores (Belsky & Israel, 2014; Schizophrenia Working Group of the Psychiatric Genomics Consortium, 2014), are a quantitative tool for aggregating a large amount of otherwise unwieldy genetic information from genome-wide association studies (GWASs), which are usually based on meta-analyses of non-probability samples of volunteers (Han et al., 2009; Houlston et al., 2010; Lindgren et al., 2009; Nalls et al., 2014; Neale et al., 2010; Sklar et al., 2011). Some researchers have recently expressed the concern that GWASs are vulnerable to selection bias for their target populations (Martin et al., 2019), motivating our current application.

For a given phenotype $p$, PGSs are generally computed as follows. First, drawing on specific GWASs focusing on that phenotype (e.g., Okbay et al., 2016), "weights" are computed for hundreds of thousands of single nucleotide polymorphisms (SNPs) from individual linear regression models. Each contributing study regresses the value for the phenotype of interest (e.g., an individual's body mass index) on the coded value for an individual SNP, typically adjusting for cohort-specific covariates. The possible values for an individual SNP range between 0 and 2, representing the expected number of alleles present on the imputed SNP based on imputation probabilities. These cohort-specific estimated coefficients are then meta-analyzed across all contributing studies. Second, the resulting coefficient from the GWAS meta-analysis, denoted by $\hat{\alpha}_{i(p)}$ for SNP $i$ and phenotype $p$, is treated as a weight in computing the PGSs in an independent sample. The study-specific PGS for phenotype $p$ for a given individual is then the linear combination of the products of the coded SNP values (denoted by $g_i$ and represented by 0, 1, or 2 copies of the risk allele) and the GWAS meta-analysis weights across all SNPs:

$$PGS_p = \sum_i \hat{\alpha}_{i(p)} g_i \qquad (1)$$

While the PGS is usually computed as in (1), various modifications have been suggested, and this is an active area of methodological research. For example, researchers can decide whether to use the correlation structure of the human genome to minimize the number of correlated variants in a score, and some researchers may employ *p*-value thresholds for identifying which weights are "important" for the computation; see Ware et al. (2017) for an in-depth discussion of these issues. Other issues with the PGS are discussed in Section 6.

Underlying the use of PGS is the assumption that it is in fact a strong correlate of the measures for the phenotype of interest; this assumption is usually checked with simple regression models for the phenotypes that include the PGS as a predictor. PGSs have been found to be useful correlates of age at onset of alcohol dependence (Kapoor et al., 2016), selected psychiatric traits (Stein et al., 2017; Wray et al., 2014), schizophrenia and bipolar disorder (International Schizophrenia Consortium, 2009; Schizophrenia Working Group of the Psychiatric Genomics Consortium, 2014), and BMI (Locke et al., 2015), among other traits. However, the genetic data used to compute PGSs are generally collected from non-probability samples (usually composed of volunteers), as in the GfG study. This raises important questions about whether the estimates of PGS-phenotype relationships are biased for the target population of interest. Recent work has suggested that the predictive ability of PGSs may be limited due to this selection bias (Martin et al., 2019). In Section 4, we address this question using the measures of selection bias developed in Section 3.

*2.2 Past-year Employment for Smartphone Users with Less Than High School Education in the National Survey of Family Growth*

A major issue in modern survey research is the potential for selection bias in samples of smartphone users, given that that smartphones are now a primary communication tool in opt-in online surveys (e.g., Revilla, 2017). Little et al. (2020) evaluated the potential non-ignorable selection bias in selected estimates of means based on self-identified smartphone users in the National Survey of Family Growth (NSFG). They assumed that smartphone users were a non-probability sample selected from a hypothetical population defined by the full NSFG sample. This subsample, however, was a larger fraction of the overall NSFG "population" than would be characteristic of most non-probability samples, given the high penetration of mobile devices in the U.S. (Blumberg and Luke, 2018). A large body of research has established positive correlations between education, current employment status, and income (e.g., Morgan and David, 1963; Muller, 2002). Research suggests that individuals with lower education may be more responsive to surveys inviting sampled persons to participate with some monetary incentive promised in return (e.g., Petrolia and Bhattacharjee, 2009; Ryu, Couper and Marans, 2005). We therefore focused this application of our proposed measures on smartphone users with less than high school education as a hypothetical non-probability sample with a smaller sampling fraction than reported by Little et al. (2020). ==We treated the full NSFG sample as the overall population, which allowed us to evaluate the ability of the measures developed in this paper to detect the selection bias associated with this hypothetical non-probability sample.==

Specifically, we sought to fit a linear regression model predicting the number of months worked in the past year as a function of gender (male / female) and age (15-18, 19-29, or 30-49), given the importance of these socio-demographic subgroups in employment research (Mandel and Semyonov, 2014). We fit this linear regression model in the "non-probability sample" defined by smartphone users with less than high school education in the NSFG ($n$ = 2,977). Our goal is to assess the selection bias in these estimated regression coefficients using auxiliary data, namely race/ethnicity (non-Hispanic White, non-Hispanic Black, Hispanic, Other), marital status (married, divorced/widowed/separated), household income (<$19,999, $20,000-$59,999, $60,000+), region of the United States (Midwest, Northeast, South, and West), current employment status (working / not working), and presence of children under the age of 16 in the household (yes, no). Unlike gender and age, these auxiliary variables were not of primary interest but were still thought to be predictive of current employment status when adjusting for gender and age. The performance of our measures of selection bias could be assessed here because we were able to compute the regression coefficients of interest for the "non-selected cases" in the remainder of the NSFG sample ($n$ = 16,823).

## 3. Models and Methods

*3.1 Basic Setting*

Assume that a non-probability sample has data $D_i = \{Y_i, Z_i, A_i, i = 1, ..., n\}$, where $n$ is the sample size, and for unit $i$, $Y_i$ is the value of an outcome variable $Y$, $Z_i$ is the value of a $p \times 1$ vector of predictor variables $Z$, interest concerns the linear regression of $Y$ on $Z$, and $A_i$ is the value of a vector of auxiliary variables $A$ not included in the regression. We assume first that $Y$ is normal; in Section 3.3 we consider extensions to probit regression where $Y$ is binary. We also assume that summary statistics, specifically means, variances and covariances of $Z$ and $A$, are either available for the population from an external source (e.g., census data or administrative records) or can be estimated from a large probability sample (e.g., the American Community Survey).

Let $S$ be a selection indicator, equal to 1 for population units included in the non-probability sample and 0 otherwise. We assume that the non-probability sample arises from a population of size $N$, where $n^{(0)} = N - n$ cases are not selected. To address potential selection bias, a model is required for the joint distribution of $Y$ and $S$ given $Z$ and $A$. Our proposed indices are based on a *pattern-mixture model* (Little, 1993; Glynn, Laird and Rubin, 1986), or PMM, which factorizes this joint distribution as

$$p(Y, S \mid Z, A, \theta, \phi) = p(Y \mid S, Z, A, \theta) p(S \mid Z, A, \phi), \qquad (1)$$

where $\theta$ and $\phi$ are unknown parameters.

Our approach therefore requires auxiliary variables $A$ that are not included in the regression model of interest, but are still predictive of $Y$ after conditioning on $Z$. This could be the case, for example, in a study focusing on descriptive survey estimates of means for different subgroups based on a non-probability sample, where the implicit underlying regression model includes main effects and interactions associated with the classification variables defining the subgroups, but omitted auxiliary variables may still be predictive of the outcome of interest (e.g., Clifford, Jewell, and Waggoner, 2015). Another example is causal inference for the effects of $Z$ on $Y$, where the variables $A$ are endogenous. In particular, in clinical trials, auxiliary variables measured post-treatment are excluded from the model for estimating the treatment effect because they incorporate effects of the treatment.

This need for auxiliary variables is limiting, so a natural question is what can be done without them. Without auxiliary variables or structural model assumptions, the data provide no information about the regression of $Y$ on $Z$ for unselected cases, and how it differs from the regression of $Y$ on $Z$ for selected cases. A competing method for factorizing the joint distribution of $Y$ and $S$ given $Z$ and $A$ is to use a *selection model*:

$$p(Y, S \mid Z, A, \gamma, \lambda) = p(Y \mid Z, A, \gamma) p(S \mid Y, Z, A, \lambda), \qquad (2)$$

where $\gamma$ and $\lambda$ are unknown parameters. The well-known selection model proposed by Heckman (1976) could be used to model the joint distribution of $Y$ and $S$ given $Z$ and $A$. However, as we discuss in more detail in Section 3.2, this approach has some serious drawbacks, including the requirement of microdata for $Z$ and $A$ for the unselected cases. Goldberger (1981, Eq. 37) presented expressions of the bias in estimated regression coefficients for the Heckman (1976) selection model, where selection is assumed to occur when a latent variable (say $L$) crosses a threshold, and $L$ is assumed to have a joint normal distribution with the outcome variable $Y$. However, properly identifying this model requires unverifiable assumptions that exclude subsets of $Z$ from the regression model for $Y$ or the model for $L$ that determines selection (see e.g. Little, 1985; Little and Rubin, 2019, Chapter 15). Our methods concern situations where such assumptions are not warranted.

*3.2 Linear Regression for Normal Outcomes*

Here and throughout the paper, we use the generic notation $(\beta_{y0 \cdot zw}^{(s)}, \beta_{yw \cdot zw}^{(s)}, \sigma_{yy \cdot zw}^{(s)})$ for respectively the intercept, regression coefficients (written as a row vector) and residual variance of the regression of $Y$ on $Z$ and $W$ given $S = s$. We focus first on the distribution $p(Y \mid S, Z, A, \theta)$ on the right side of the PMM in (1). We suppose that

$$E(Y \mid S = 1, Z, A, \theta) = \beta_{y0 \cdot za}^{(1)} + \beta_{yz \cdot za}^{(1)} Z + \beta_{ya \cdot za}^{(1)} A,$$

where $X = \left(\beta_{ya \cdot za}^{(1)} A\right)$ denotes the best predictor of $Y$ in the non-probability sample ($S = 1$) after conditioning on $Z$, and $\theta$ represents all of the parameters defining this distribution. Let $V$ denote a vector of variables in $A$ that is orthogonal to $X$ after conditioning on $Z$; that is, in a linear regression model for $X$ on $Z$ and $V$, $\beta_{xv \cdot zv}^{(1)} = 0$. The variables in $V$ will therefore generally be those variables in $A$ that are *not* predictive of $Y$ when conditioning on $Z$. If $A$ is a single variable, then $X = A$ and $V$ is null.

We now assume the following model for the distribution of $X$ and $Y$ given $S$, $Z$, and $V$:

$$\left(\binom{X}{Y} \mid S = s, Z, V, \theta\right) \sim N_2 \left( \begin{pmatrix} \beta_{x0 \cdot zv}^{(s)} + \beta_{xz \cdot zv}^{(s)} Z + \beta_{xv \cdot zv}^{(s)} V \\ \beta_{y0 \cdot zv}^{(s)} + \beta_{yz \cdot zv}^{(s)} Z + \beta_{yv \cdot zv}^{(s)} V \end{pmatrix}, \begin{pmatrix} \sigma_{xx \cdot zv}^{(s)} & \sigma_{xy \cdot zv}^{(s)} \\ \sigma_{xy \cdot zv}^{(s)} & \sigma_{yy \cdot zv}^{(s)} \end{pmatrix} \right), \qquad (3)$$

where $N_2(\mu, \Sigma)$ denotes the bivariate normal distribution with mean $\mu$ and covariance matrix $\Sigma$. We make the following additional assumptions about this model:

a) $E(Y \mid S = 0, Z, X, V, \theta) = \beta_{y0 \cdot zxv}^{(0)} + \beta_{yz \cdot zxv}^{(0)} Z + \lambda^{(0)} X$; that is, $X = \left(\beta_{ya \cdot za}^{(1)} A\right)$ is the best predictor of $Y$, after conditioning on $Z$, for both selected and non-selected ($S = 0$) cases; and

b) $V$ is orthogonal to $X$ given $Z$ for non-selected as well as selected cases; that is, $\beta_{xv \cdot zv}^{(s)} = 0$ for $S = 1$ and $S = 0$. This assumption can be checked given microdata on the non-selected cases.

Under assumptions (a) and (b), $\beta_{yv \cdot xzv}^{(s)} = \beta_{yv \cdot zv}^{(s)} - \sigma_{xy \cdot zv}^{(s)} \beta_{xv \cdot zv}^{(s)} / \sigma_{xx \cdot zv}^{(s)} = \beta_{yv \cdot zv}^{(s)} = 0$ for both $S = 1$ and $S = 0$, so the distribution in (3) reduces to

$$\left(\binom{X}{Y} \mid S = s, Z, V, \theta\right) \sim N_2 \left( \begin{pmatrix} \beta_{x0 \cdot zv}^{(s)} + \beta_{xz \cdot zv}^{(s)} Z \\ \beta_{y0 \cdot zv}^{(s)} + \beta_{yz \cdot zv}^{(s)} Z \end{pmatrix}, \begin{pmatrix} \sigma_{xx \cdot zv}^{(s)} & \sigma_{xy \cdot zv}^{(s)} \\ \sigma_{xy \cdot zv}^{(s)} & \sigma_{yy \cdot zv}^{(s)} \end{pmatrix} \right). \qquad (4a)$$

Because $X$ and $Y$ are independent of $V$ given $Z$ and $S$, we can simplify the notation in (4a) by dropping the subscript $v$ in the parameters, resulting in the equivalent model

$$\left(\binom{X}{Y}\mid S=s,Z,V,\theta\right) \sim N_2\left(\begin{pmatrix}\beta_{x0\cdot z}^{(s)}+\beta_{xz\cdot z}^{(s)}Z\\ \beta_{y0\cdot z}^{(s)}+\beta_{yz\cdot z}^{(s)}Z\end{pmatrix},\begin{pmatrix}\sigma_{xx\cdot z}^{(s)} & \sigma_{xy\cdot z}^{(s)}\\ \sigma_{xy\cdot z}^{(s)} & \sigma_{yy\cdot z}^{(s)}\end{pmatrix}\right). \qquad (4b)$$

We focus next on the distribution of $S$ given $Z$ and $A$ in (1), $p(S\mid Z,A,\phi)$. Following Little (1994) and Andridge and Little (2011), this distribution is left unspecified, and the unidentified parameters in (4b) are identified by assuming that given $X$, $Y$, $Z$, and $V$, selection depends on $X$ and $Y$ only through a known linear combination of these variables. Specifically, we assume that

$$\Pr(S=1\mid X,Y,Z,V,\phi) = g(Y_\phi, Z, V), \text{ with } Y_\phi = \phi Y + (1-\phi)X^*, \qquad (5)$$

where $g()$ is an arbitrary unknown function, and $X^* = X\sqrt{\sigma_{yy\cdot z}^{(1)}/\sigma_{xx\cdot z}^{(1)}}$ is $X$ rescaled to have the same residual variance as $Y$ after conditioning on $Z$. We call $X^*$ the *scaled auxiliary proxy* for $Y$; the transformation from $X$ to $X^*$ simplifies the interpretation of $\phi$ by putting $Y$ and $X^*$ on the same scale.

We noted above that if $A$ is univariate, then $X=A$ and $V$ is null, in which case $A$ is the best predictor because it is the *only* predictor of $Y$ after conditioning on $Z$. When $A$ consists of two or more variables, the formulation in (4b) and (5) is actually a refinement of the original formulation of the proxy PMM in Andridge and Little (2011), which did not explicitly include $V$ as part of the model, and assumed that

$$\Pr(S=1\mid X,Y,Z) = g(Y_\phi, Z), \text{ with } Y_\phi = \phi Y + (1-\phi)X^*.$$

The refinement is that, with the additional assumptions (a) and (b) above, the selection propensity is allowed to depend on $V$ as well as $Y_\phi$, which is a much weaker and more realistic assumption. For example, if $Y$ is income and $A$ = (age, sex, occupation, education, race), then selection can depend on *all* the variables in $A$ (including those in $V$ that are orthogonal to $X$ when conditioning on $Z$), rather than just the particular linear combination of the variables in $A$ that best predicts income ($X$). We note that the combination of variables in $A$ that best predicts *selection* would not necessarily be the same as the combination that best predicts *income*.

The assumption (5) identifies the model (4b) when $\phi$ is known. In practice, $\phi$ is unknown; because $X$ is an auxiliary proxy for $Y$, we assume that $\phi$ is positive, that is, $0 \leq \phi \leq 1$. When $\phi = 0$, selection depends only on the observed variables $Z$, $X$, and $V$, and hence is "selection at random," a condition analogous to missing at random in Rubin (1976). At the other extreme when $\phi = 1$, selection depends only on $Y$, $Z$, and $V$. The parameter $\phi$ is therefore a measure of the "degree of non-random selection," after conditioning on $Z$, $X$, and $V$, and no information is available on $\phi$ in the data.

*Selection Mechanism 1: $\phi = 1$.* Consider first the setting where $\phi = 1$ in (5). This implies that selection depends only on $Y$, $Z$, and $V$, and therefore the regression of $X$ on $Y$, $Z$, and $V$ is the same for both patterns defined by $S$. Hence, under assumption (b) above, we have

$$\beta_{x0\cdot yz}^{(1)} = \beta_{x0\cdot yz}^{(0)}, \beta_{xy\cdot yz}^{(1)} = \beta_{xy\cdot yz}^{(0)}, \text{ and } \beta_{xz\cdot yz}^{(1)} = \beta_{xz\cdot yz}^{(0)}. \qquad (6)$$

Maximum likelihood (ML) estimates (or Bayesian posterior draws) of these parameters can therefore be obtained from the regression of the auxiliary proxy $X$ on $Y$ and $Z$ for the non-probability sample ($S = 1$). ML estimates (or draws) of the parameters $\left(\beta_{x0 \cdot z}^{(s)}, \beta_{xz \cdot z}^{(s)}, \sigma_{xx \cdot z}^{(s)}\right)$ from the regression of $X$ on $Z$ can also be obtained from regression models fitted to each respective pattern ($S = 0,1$). We can use the microdata from the non-probability sample to obtain estimates of these parameters for the selected cases ($S = 1$).

In the absence of microdata for the non-selected cases ($S = 0$), we only need means, variances, and covariances of $Z$ and $A$ for these cases to compute ML estimates (or draws) of the parameters $\left(\beta_{x0 \cdot z}^{(0)}, \beta_{xz \cdot z}^{(0)}, \sigma_{xx \cdot z}^{(0)}\right)$ from the regression of $X$ on $Z$. Assuming a negligible sampling fraction for the non-probability sample, these estimated means, variances, and covariances could be computed by performing appropriately-weighted analyses of the survey data from a large probability sample (e.g., the Health and Retirement Study; https://hrs.isr.umich.edu/about) that collected measures of all variables in the vectors $Z$ and $A$ (including appropriate dummy variables for the levels of any categorical variables).

If the sampling fraction for the non-probability sample cannot be considered negligible (given the relative sizes of the non-probability sample and the target population), one could employ a variation of the quasi-randomization method outlined in Elliott and Valliant (2017), using a reference probability sample to model the probability of *not* being included in the non-probability sample. Pseudo-weights defined by the inverses of these predicted probabilities could be applied to cases from the probability sample to compute estimates of the required means, variances, and covariances for the non-selected cases.

Given valid, design-consistent estimates of the means, variances, and covariances of $Z$ and $A$ for the non-selected cases, estimates (or draws) of the parameters $\left(\beta_{x0 \cdot z}^{(0)}, \beta_{xz \cdot z}^{(0)}, \sigma_{xx \cdot z}^{(0)}\right)$ for the non-selected cases would be computed using the following three steps, for both the setting currently being discussed (where $\phi = 1$) and all other possible values of $\phi$. First, we define the variance-covariance matrix for $X$ and $Z$ for the non-selected cases as follows, where $\Sigma_{aa}^{(0)}$, $\Sigma_{az}^{(0)}$, and $\Sigma_{zz}^{(0)}$ refer to the estimated variance-covariance matrix for the $A$ variables, the estimated covariances of the $A$ and $Z$ variables, and the estimated variance-covariance matrix for the $Z$ variables, respectively, computed for the non-selected cases ($S = 0$) based on the data from the aforementioned large probability sample:

$$\operatorname{var}(X, Z)^{(0)} = \begin{bmatrix} \beta_{ya \cdot za}^{(1)} \Sigma_{aa}^{(0)} \beta_{ya \cdot za}^{(1)T} & \beta_{ya \cdot za}^{(1)} \Sigma_{az}^{(0)} \\ \beta_{ya \cdot za}^{(1)} \Sigma_{az}^{(0)} & \Sigma_{zz}^{(0)} \end{bmatrix} \equiv \begin{bmatrix} \sigma_{xx}^{(0)} & \Sigma_{xz}^{(0)} \\ \Sigma_{xz}^{(0)} & \Sigma_{zz}^{(0)} \end{bmatrix}. \quad (7)$$

Second, we compute $\beta_{xz \cdot z}^{(0)} = \left[\Sigma_{zz}^{(0)}\right]^{-1} \Sigma_{xz}^{(0)}$ and $\beta_{x0 \cdot z}^{(0)} = \overline{X}^{(0)} - \beta_{xz \cdot z}^{(0)} \overline{Z}^{(0)}$, where the vector of means $\overline{X}^{(0)}$ is computed using 1) the estimated coefficients for the $A$ variables from the regression of $Y$ on $Z$ and $A$ for the selected cases, and 2) the estimated means of the $A$ variables for the non-selected cases. Third, we compute

$$\sigma_{xx \cdot z}^{(0)} = \left[\frac{n^{(0)}}{n^{(0)} - p}\right] \left[\sigma_{xx}^{(0)} - \Sigma_{xz}^{(0)} \left[\Sigma_{zz}^{(0)}\right]^{-1} \Sigma_{xz}^{(0)}\right],$$

where $n^{(0)}$ denotes the number of non-selected cases in the population and $p$ is the number of predictors in $Z$. Making the same assumption about the negligible sampling fraction associated with the non-probability sample discussed above, we note that $n^{(0)}/(n^{(0)}-p)$ will generally be quite close to 1 in non-probability samples with $n^{(0)} \gg p$.

We can now express the unidentified parameters of the linear regression of $Y$ on $Z$ for $S = 0$ in terms of the identified parameters above. The intercept of the linear regression of $Y$ on $Z$ for $S = 0$ can be written as

$$\beta_{y0 \cdot z}^{(0)} = \frac{\beta_{x0 \cdot z}^{(0)} - \beta_{x0 \cdot yz}^{(0)}}{\beta_{xy \cdot yz}^{(0)}} =_{\text{[by (6)]}} \frac{\beta_{x0 \cdot z}^{(0)} - \beta_{x0 \cdot yz}^{(1)}}{\beta_{xy \cdot yz}^{(1)}} = \frac{\beta_{x0 \cdot z}^{(0)} - \left(\beta_{x0 \cdot z}^{(1)} - \beta_{xy \cdot yz}^{(1)}\beta_{y0 \cdot z}^{(1)}\right)}{\beta_{xy \cdot yz}^{(1)}}, \tag{8}$$

and hence $\beta_{y0 \cdot z}^{(0)} = \beta_{y0 \cdot z}^{(1)} + \left(\beta_{x0 \cdot z}^{(0)} - \beta_{x0 \cdot z}^{(1)}\right)/\beta_{xy \cdot yz}^{(1)}$. Similarly, for the slope of $Z$ and the residual variance of the linear regression of $Y$ on $Z$, we have:

$$\beta_{yz \cdot z}^{(0)} = \beta_{yz \cdot z}^{(1)} + \frac{\beta_{xz \cdot z}^{(0)} - \beta_{xz \cdot z}^{(1)}}{\beta_{xy \cdot yz}^{(1)}}, \sigma_{yy \cdot z}^{(0)} = \sigma_{yy \cdot z}^{(1)} + \frac{\sigma_{xx \cdot z}^{(0)} - \sigma_{xx \cdot z}^{(1)}}{\left(\beta_{xy \cdot yz}^{(1)}\right)^2}. \tag{9}$$

ML estimates (or draws) of these parameters can be obtained by substituting the ML estimates (or draws) of the identified parameters on the right-hand sides of the expressions in (8) and (9).

*Selection Mechanism 2:* $0 \leq \phi < 1$. Following Andridge and Little (2011), the transformation $Y_\phi = \phi Y + (1-\phi)X^*$ introduced in (5) for other values of $\phi$ less than 1 yields

$$\beta_{y0 \cdot z}^{(0)} = \beta_{y0 \cdot z}^{(1)} + \left(\frac{\phi + (1-\phi)\rho_{xy \cdot z}^{(1)}}{\phi \rho_{xy \cdot z}^{(1)} + (1-\phi)}\right)\sqrt{\frac{\sigma_{yy \cdot z}^{(1)}}{\sigma_{xx \cdot z}^{(1)}}}\left(\beta_{x0 \cdot z}^{(0)} - \beta_{x0 \cdot z}^{(1)}\right), \tag{10}$$

where the correlation of $X$ and $Y$ conditional on $Z$ for the selected cases ($S = 1$) is $\rho_{xy \cdot z}^{(1)} = \sigma_{xy \cdot z}^{(1)}/\sqrt{\sigma_{xx \cdot z}^{(1)}\sigma_{yy \cdot z}^{(1)}}$. We also have

$$\beta_{yz \cdot z}^{(0)} = \beta_{yz \cdot z}^{(1)} + \left(\frac{\phi + (1-\phi)\rho_{xy \cdot z}^{(1)}}{\phi \rho_{xy \cdot z}^{(1)} + (1-\phi)}\right)\sqrt{\frac{\sigma_{yy \cdot z}^{(1)}}{\sigma_{xx \cdot z}^{(1)}}}\left(\beta_{xz \cdot z}^{(0)} - \beta_{xz \cdot z}^{(1)}\right) \text{ and}$$

$$\sigma_{yy \cdot z}^{(0)} = \sigma_{yy \cdot z}^{(1)} + \left(\frac{\phi + (1-\phi)\rho_{xy \cdot z}^{(1)}}{\phi \rho_{xy \cdot z}^{(1)} + (1-\phi)}\right)^2 \left(\frac{\sigma_{yy \cdot z}^{(1)}}{\sigma_{xx \cdot z}^{(1)}}\right)\left(\sigma_{xx \cdot z}^{(0)} - \sigma_{xx \cdot z}^{(1)}\right). \tag{11}$$

As before, ML estimates (or draws) of these parameters are obtained by substituting ML estimates (or draws) of the identified parameters into these expressions.

We propose the differences between the ML estimates (or Bayesian posterior draws) of the parameters of the regression of $Y$ on $Z$ for the selected and non-selected cases as a *Measure of Unadjusted Bias* for the regression coefficients as compared to the *Non-Selected* cases (MUBNS). After manipulating the general result in (10), the resulting MUBNS for the intercept can be written as

$$\text{MUBNS}_0(\phi) = \beta_{y0 \cdot z}^{(1)} - \beta_{y0 \cdot z}^{(0)} = \left(\frac{\phi + (1-\phi)\hat{\rho}_{xy \cdot z}^{(1)}}{\phi \hat{\rho}_{xy \cdot z}^{(1)} + (1-\phi)}\right)\sqrt{\frac{\hat{\sigma}_{yy \cdot z}^{(1)}}{\hat{\sigma}_{xx \cdot z}^{(1)}}}\left(\hat{\beta}_{x0 \cdot z}^{(1)} - \hat{\beta}_{x0 \cdot z}^{(0)}\right) \tag{12}$$

and similarly, following (11), the MUBNS indices for the slopes can be written as

$$\text{MUBNS}_z(\phi) = \beta_{yz \cdot z}^{(1)} - \beta_{yz \cdot z}^{(0)} = \left(\frac{\phi + (1-\phi)\hat{\rho}_{xy \cdot z}^{(1)}}{\phi \hat{\rho}_{xy \cdot z}^{(1)} + (1-\phi)}\right)\sqrt{\frac{\hat{\sigma}_{yy \cdot z}^{(1)}}{\hat{\sigma}_{xx \cdot z}^{(1)}}}\left(\hat{\beta}_{xz \cdot z}^{(1)} - \hat{\beta}_{xz \cdot z}^{(0)}\right). \tag{13}$$

If the selection fraction is small (as is the case with most non-probability samples), the differences defining the MUBNS indices in (12) and (13) essentially capture the bias in the regression coefficients estimated from the selected cases relative to the regression coefficients based on the entire population. By the law of total probability, we know that

$$E(Y|Z, S=1) - E(Y|Z) = [E(Y|Z, S=1) - E(Y|Z, S=0)] \times \Pr(S=0|Z). \tag{14}$$

The pattern mixture model specified in (4b) and (5) provides a comparison of the regression coefficients for $S = 1$ and $S = 0$, as in the first term on the right-hand side of (14). For a comparison with the regression coefficients for the whole population [the *entire* right-hand side of (14)], the impact of the difference in coefficients at a particular value of $Z$ depends on the non-selection rate $\Pr(S = 0|Z)$ for that value of $Z$. We note that the relative impact of selection on coefficients for different $Z$ variables does not depend on $Z$, that is, $\Pr(S = 0|Z)$ is a constant factor in this comparison. If the overall selection rate for the non-probability sample is non-negligible and is known or can be estimated, we propose a *Measure of Unadjusted Bias* (MUB) for the selected cases that compares the coefficients to those for the entire population, by multiplying the MUBNS indices by the overall non-selection rate:

$$\text{MUB}_0(\phi) = \text{MUBNS}_0(\phi) \times \Pr(S=0) \text{ and } \text{MUB}_z(\phi) = \text{MUBNS}_z(\phi) \times \Pr(S=0). \tag{15}$$

We make the following five remarks about the indices proposed in (12), (13), and (15):
1. The MUB indices in (15) could be used to make inferences about the regression coefficients after adjusting for the selection bias, simply by subtracting the MUB indices from the estimates (or draws) of the coefficients for the selected sample.
2. In the case where the regression model of interest only includes an intercept (i.e., $Z$ does not exist), the MUB index defined in (15) equals the unstandardized MUB index presented in Little et al. (2020) for means of continuous variables.
3. We recommend defining posterior distributions for the proposed indices by performing a fully Bayesian analysis with a prior distribution on $\phi$, as described by Little et al. (2020) and Andridge et al. (2019). One can then use credible intervals for the indices defined in (12), (13), and (15) to make inference about the selection bias. We consider this Bayesian approach, outlined in detail in the online supplementary materials, in our simulation study and our applications. Importantly, this Bayesian approach allows us to fully account for the uncertainty in all input estimates required for the proposed indices (e.g., the estimated regression coefficients used to form the auxiliary proxy $X$).
4. Little et al. (2020) and Andridge et al. (2019) also note the importance of having a substantial correlation between $X$ and $Y$, which in our regression framework corresponds to having a substantial value of $\rho_{xy \cdot z}^{(1)}$, for these indices to be effective indicators of selection bias.
5. An alternative approach is to apply the Heckman (1976) selection model to the distribution of $Y$ and $S$ given $Z$ and $A$. However, our approach offers a number of advantages. First, our measure of selection bias is simpler and easier to interpret than the corresponding expressions in the selection model approach; see Eq. (37) in Goldberger (1981). Second, our method is computationally simpler, because the selection model involves an iterative fitting algorithm for each value of a sensitivity parameter. Third, fitting the selection model requires microdata on $Z$ and $A$ for the non-selected cases, which is often highly unrealistic. Our proposed measures only require summary statistics for $Z$ and $A$ for the non-selected cases.

*3.3 Probit Regression for Binary Outcomes*

We now extend the indices of selection bias developed above to the case of binary outcome variables. Recent work by Maity et al. (2020) examined alternative methods of correcting for the selection bias in estimated logistic regression model coefficients when missingness on the binary outcome is nonignorable. Similar to the important limitation of Heckman's selection model discussed above, all of the methods evaluated by Maity and colleagues require microdata on the predictors of interest and any auxiliary variables for the (non-selected) cases with missing values on the binary outcome. In this section, we describe how our proposed indices, which are not subject to this limitation, can be extended to the case of probit regression models for binary outcomes.

In the case where $Y$ is a binary variable, taking on values of 1 and 0, and the parameters of interest are the coefficients in a probit regression model of $Y$ on $Z$, the pattern-mixture model (4b) and (5) can be applied to an underlying latent standard normal variable $U$ that gives rise to $Y$ (where $Y = 1$ if $U > 0$). Specifically, under the same assumptions outlined above, we have

$$\left( \begin{pmatrix} X \\ U \end{pmatrix} \mid Z, V, S \right) \sim N_2 \left( \begin{pmatrix} \beta_{x0\cdot z}^{(s)} + \beta_{xz\cdot z}^{(s)t} Z \\ \beta_{u0\cdot z}^{(s)} + \beta_{uz\cdot z}^{(s)t} Z \end{pmatrix}, \begin{pmatrix} \sigma_{xx\cdot z}^{(s)} & \sigma_{xu\cdot z}^{(s)} \\ \sigma_{xu\cdot z}^{(s)} & \sigma_{uu\cdot z}^{(s)} \end{pmatrix} \right), \quad (16)$$

where the conditional probability of selection is defined as in (5), replacing $Y$ with $U$. Following Andridge and Little (2020), the conditional variance of $U$ (given $Z$ and $A$) in the model initially fitted to create the auxiliary proxy $X$ is fixed at 1 (i.e., $\sigma_{uu\cdot za}^{(1)} = 1$) since the mean and variance of the latent variable cannot both be estimated. Given the model in (16), we arrive at the same indices of selection bias defined in (12), (13), and (15).

The key distinction in this case is that the latent standard normal variable $U$ is not directly observed. We therefore propose an update of our fully Bayesian approach for this binary case, using a Gibbs sampler to repeatedly draw values of the latent variable $U$ based on the observed data (and subsequently draw values of the indices of selection bias for the coefficients in the probit regression model of interest). We add the following steps to the beginning of the Bayesian algorithm outlined in the online supplementary materials:

1. Fit a probit regression model to the observed binary outcome $Y$ for the selected cases ($S = 1$), including the $Z$ and $A$ variables as predictors, yielding the coefficients $\hat{\beta}_{yz\cdot za}^{(1)}$ and $\hat{\beta}_{ya\cdot za}^{(1)}$. Let $\hat{\beta} = \left( \hat{\beta}_{yz\cdot za}^{(1)}, \hat{\beta}_{ya\cdot za}^{(1)} \right)$, and compute the auxiliary proxy $X_i^{(1)} = \hat{\beta}_{ya\cdot za}^{(1)} A_i$.

2. For each selected case $i$, we draw $U_i^{(d)(1)} \sim N(\hat{\beta}_{yz\cdot za}^{(1)} Z_i + X_i^{(1)}, 1 \mid y_i^{(1)}, \hat{\beta}, S = 1)$, where the draws are truncated below at zero if $y_i = 1$ and above at zero if $y_i = 0$.

3. Given $Z_i$, $A_i$, and $U_i^{(d)(1)}$ for all $i$ with $S = 1$, draw $\hat{\beta}^{(d)}$ from a multivariate normal distribution with mean defined by the coefficients from a linear model of $U^{(d)(1)}$ on $Z$ and $A$ with the residual variance constrained to be 1, and variance-covariance matrix defined by the estimated variance and covariances of the coefficients based on this model.

4. Given $\hat{\beta}^{(d)}$, compute a draw of the auxiliary proxy $X_i^{(d)(1)} = \hat{\beta}_{ya\cdot za}^{(d)(1)} A_i$.

Given the results in Steps 2 through 4 above, we can now proceed with Step 2 of the Bayesian approach outlined in the supplementary materials for the normal case (computing the relevant statistics for the non-selected cases as described earlier). However, we note that in the linear regression of the draws of $U$ (from Step 2 above) on $Z$, which is an important part of this second step of the Bayesian approach, the conditional variance $\sigma_{uu \cdot z}^{(1)}$ in (16) will be greater than 1, given that we are no longer conditioning on $A$. Because we require that this conditional variance be equal to 1 in the probit regression model of interest, we need to scale the estimated coefficients in this linear regression model appropriately, using the estimate of $\sigma_{uu \cdot z}^{(1)}$ from the regression. We therefore introduce this modification to Step 2 of the Bayesian approach, and divide each estimated coefficient in the linear regression by a draw of $\sigma_{uu \cdot z}^{(1)}$ before proceeding with the subsequent steps in the supplementary materials.

After obtaining a posterior draw of the indices of selection bias, we return to Step 2 above to repeat the process and obtain additional posterior draws of the indices. This process can be repeated several times to simulate draws from the posterior distribution for the indices of selection bias for the probit regression model coefficients, accounting for the uncertainty in all input estimates. We have included R functions implementing the computation of the indices described in Sections 3.2 and 3.3, along with annotated examples, in the GitHub repository https://github.com/bradytwest/IndicesOfNISB.

## 4. Simulation Study

*4.1 Design of the Simulation Study*

We assess the effectiveness of the proposed MUBNS indices via a simulation study; the study also serves as an assessment of the effectiveness of the MUB indices when the selection rate is known or can be estimated. Let $Y$ be the outcome variable of interest, let $Z_1$ and $Z_2$ be the predictor variables of interest in the target linear regression model, and let $A$ be an auxiliary variable, with population-level summary statistics available for the $Z_1$, $Z_2$ and $A$. We repeatedly generate populations of size $N = 10,000$ units from the following superpopulation model:

$$\begin{pmatrix} Y \\ Z_1 \\ Z_2 \\ A \end{pmatrix} \sim N \left( \begin{pmatrix} 10 \\ 0 \\ 0 \\ 0 \end{pmatrix}, \begin{pmatrix} 4 & 2\rho_{y1} & 2\rho_{y2} & \sigma_{ya} \\ 2\rho_{y1} & 1 & 0 & \rho_{1a} \\ 2\rho_{y2} & 0 & 1 & 0 \\ \sigma_{ya} & \rho_{1a} & 0 & 1 \end{pmatrix} \right) \quad (17)$$

Note that the predictor variables of interest are independent, with $Z_1$ correlated with $A$, and $Z_2$ uncorrelated with $A$.

The correlations of $Y$ with $Z_1$ and $Z_2$, say $\rho_{y1}$ and $\rho_{y2}$, are set to 0.2 (low), 0.4 (medium) or 0.6 (high), and the correlation of $Y$ and $A$ given $Z_1$ and $Z_2$ is set to 0.2, 0.5 or 0.8. We then determine the values of $\sigma_{ya}$ given these values. The correlation $\rho_{1a}$ between $Z_1$ and $A$ is set to 0.2, 0.4 or 0.6. The combinations of these parameter choices result in $3 \times 3 \times 3 \times 3 = 81$ possible population distributions.

The probability that a unit from a simulated population is included in (or selected for) a non-probability sample is determined by the following selection model:
$$\text{logit}\big(P(S = 1 | Y, Z_1, Z_2, A)\big) = \gamma_0 + \gamma_y Y + \gamma_{Z1} Z_1 + \gamma_{Z2} Z_2 + \gamma_a A, \quad (18)$$
where $S$ is the selection indicator (1 = selected, 0 = not selected). The selection model in (18) would be unknown to the analyst. Values of the parameters in (18) are defined as follows:
- $\gamma_y = \{0, \ln(1.1), \ln(2)\}$; here, $\gamma_y = 0$ implies Selection At Random
- $\gamma_{Z1} = \{\ln(1.1), \ln(2)\}$

- $\gamma_{Z2} = \{\ln(1.1), \ln(2)\}$
- $\gamma_a = \{\ln(1.1), \ln(2)\}$

These values represent either no effects on selection (*Y* only; OR=1), small effects on selection (all variables; OR=1.1), or strong effects on selection (all variables; OR=2). The various combinations of these parameters result in $3 \times 2 \times 2 \times 2 = 24$ possible selection mechanisms. For each choice, we set $\gamma_0$ to the value that results in a 5% selection fraction for the population.

For each unit in the population, we draw a UNIFORM(0,1) random number, and set $S = 1$ for that unit if the draw is less than the probability of selection based on (18), and $S = 0$ otherwise. We note that the simulated data are generated using a selection model, rather than the pattern-mixture model in (4b) and (5), so the model in (4b) and (5) does not hold exactly for the simulated data sets. The complete simulation experiment therefore features $81 \times 24 = 1,944$ combinations of data generation model and selection mechanism. For each of these combinations, we repeated the process of simulating a population of size $N = 10,000$ units and applying the specific selection mechanism 1,000 times. The simulations were programmed in R, and the simulation code is available at https://github.com/bradytwest/IndicesOfNISB.

The intercept and slopes from the linear regression of *Y* on $Z_1$ and $Z_2$ were the parameters of interest, and thus for each simulated non-probability sample we computed the values of the proposed MUBNS indices at $\phi = \{0, 0.5, 1\}$ for each of these parameters. We are not aware of any competing measures of selection bias that do not require microdata for the non-selected cases, so our assessment is limited to the MUBNS indices. The auxiliary proxy variable *X* was computed as the product of 1) the estimated coefficient for *A* from the linear regression of *Y* on $Z_1$, $Z_2$ and *A* and 2) the value of *A*. We then compared the computed MUBNS indices to the true estimated differences between the regression parameters for the selected and non-selected cases, which were available in each simulated dataset. For our first set of evaluations, we plotted the true values of the differences between coefficients against the computed values of the indices and compared the resulting relationships to a line representing a perfect 1:1 relationship. For our second set of evaluations, we examined side-by-side boxplots showing the distributions of Spearman correlations of the true differences between the coefficients for the selected cases and the population coefficients (i.e., the bias in the coefficients for the selected cases) and the MUBNS indices as a function of $\phi$.

Finally, following Little et al. (2020), we computed the percentage of simulated scenarios where intervals defined by [MUBNS(0), MUBNS(1)] (denoted by "MLE") covered the true difference in the coefficients. We also evaluated the coverage properties and median widths of 95% credible intervals for MUBNS (based on the 2.5 and 97.5 percentiles of the distribution of posterior draws of MUBNS) following the fully Bayesian approach outlined in the online supplementary materials. We considered two potential approaches for drawing values of $\phi$ when following the fully Bayesian approach: random draws from a UNIFORM(0,1) distribution ("Bayes-Uniform") and random draws from a discrete distribution where values of 0, 0.5, and 1.0 have equal probability ("Bayes-Discrete").

*4.2 Simulation Study Results*

Figure 1 presents results from all simulated scenarios and illustrates the associations between the median value of MUBNS across the 1,000 simulations and the true differences for the $Z_1$ coefficient in the model of interest (very similar results were found for the other two coefficients). When *Y* is independent of the probability of selection (row 1 of Figure 1), MUBNS(0) correlates perfectly with the difference, as expected, and the MUBNS(0.5) and MUBNS(1) indices do not perform as well. Notably, the performance of MUBNS(0.5) and MUBNS(1) improves with stronger conditional correlations between *A* and *Y* in these and all other scenarios, i.e., these estimates are closer to the true difference. In the two non-ignorable scenarios (rows 2 and 3 of Figure 1), the performance of MUBNS(0) becomes weaker as the dependence of selection on *Y* becomes stronger (going down the rows of Figure 1), and we see that MUBNS(0.5) and MUBNS(1) tend to be closer to the actual differences. MUBNS(0.5) tends to work well in most scenarios, supporting the idea of computing this index as a starting point for assessing potential bias (consistent with the recommendations of Little et al., 2020). The poor performance of MUBNS(1) illustrated in the first two panels of the third row arises when *A* has a stronger association with selection and the conditional correlation between *A* and *Y* is weaker.

Figure 2 presents the distributions of the Spearman correlations between the MUBNS index values and the true biases under the different scenarios. The clear story that emerges from this set of results is the importance of the conditional correlation between *A* and *Y* for maximizing the Spearman correlation between the MUBNS index and the true difference in the coefficients for selected and non-selected cases. The MUBNS indices correlate reasonably well with the true difference (bias) when $Cor(Y,A|Z_1,Z_2)$ is high but do worse as $Cor(Y,A|Z_1,Z_2)$ decreases. The correlations between MUBNS and the true bias vary little across all possible scenarios considered in one of these 18 panels, with most of the uncertainty emerging for the intercept and when the conditional correlation is 0.2. Since each panel contains results that pool across values of $\{\gamma_1, \gamma_2\}$, which are the log-odds of selection for $Z_1$ and $Z_2$, we can conclude that how strongly $Z_1$ and $Z_2$ are associated with selection does not have much impact on the performance of the MUBNS indices. Similarly, each of the 18 panels combines results across all values of $\{\rho_{y1}, \rho_{y2}, \rho_{1a}\}$, suggesting that the correlations of *Y* and *A* with $Z_1$ and $Z_2$ are not as influential as the conditional correlation between *Y* and *A* given $Z_1$ and $Z_2$.

Figure 3 presents empirical distributions of the rates at which the proposed [MUBNS(0), MUBNS(1)] intervals (based on the MLEs) and the Bayesian intervals ("Bayes-Uniform" and "Bayes-Discrete") cover the true differences in the coefficients across the different scenarios. The coverage of the proposed MLE-based interval improves when the dependence of selection on the dependent variable *Y* becomes stronger, and especially when selection depends more on the auxiliary proxy *A*. Notably, the coverage rates *decrease* when the auxiliary proxy *A* has a stronger conditional association with *Y*. This counter-intuitive result emerges because 1) the MLE-based intervals become narrower in the presence of more informative auxiliary data (e.g., Figure 4), and 2) for nearly-ignorable selection mechanisms, the true MUBNS is close to the lower bound of the MLE interval, which is MUBNS(0), and computing MUBNS(1) as an upper bound is inconsistent with the actual mechanism.

Figure 3 also shows that the Bayesian intervals have improved coverage of the actual differences in the coefficients relative to the MLE-based intervals in nearly all scenarios, with coverage improving given stronger auxiliary proxies and declining only for the intercept when selection depends on *A* and not *Y* (the first two columns). We do note that in the specific scenario where $\rho_{y1}$ is 0.2 ($Z_1$ is weakly associated with *Y*), $\rho_{y2}$ is 0.6 ($Z_2$ is strongly associated with *Y*), the correlation between *Y* and *A* given $Z_1$ and $Z_2$ is 0.8 (we have access to a strong proxy / auxiliary information), and $Z_1$ has a strong correlation with *A* (0.6), the Bayesian intervals tend to over-cover the difference in the $Z_1$ coefficients (at least 0.99 coverage) across all missingness mechanisms. This high coverage needs to be weighed against the width of the resulting credible intervals, which we consider next.

To examine whether the good coverage of the Bayesian intervals in Figure 3 is simply arising from wide intervals, Figure 4 presents empirical distributions of the median widths of the intervals under the different scenarios. For context, the empirical ranges of median MUBNS values for the three coefficients across the different simulation scenarios were (0.02, 19.52), (-2.21, 0.01), and (-1.87, 0.02), respectively. If one were to consider "typical" MUBNS values of 3, -1, and -1 for each coefficient, an excessively wide 95% credible interval would have a width at least 33% larger than the estimate itself, meaning that widths of 4, 1.33, and 1.33 would be considered "excessive" for these typical MUBNS values. Figure 4 shows that the median widths of the MLE-based and Bayesian credible intervals are generally quite reasonable across most of the scenarios (including the case of low conditional correlations of $A$ with $Y$). The credible intervals based on the discrete prior for $\phi$ tend to become slightly wider in the presence of less informative auxiliary information.

When the conditional correlation of the auxiliary proxy $A$ with $Y$ becomes 0.5 or higher, the Bayesian credible intervals generally achieve good coverage of the actual differences in coefficients with acceptably narrow intervals for most scenarios. The results in Figure 4 are for an intermediate association of $A$ with selection; similar patterns were found for other scenarios. Collectively, the results of our simulation study provide support for the fully Bayesian approach with a UNIFORM(0,1) prior for $\phi$.

## 5. Applications

*5.1 Polygenic Score-Phenotype Relationships in the Genes for Good Study*

For the GfG application described in Section 2.1, we first assess selection bias for several PGS-Phenotype relationships computed using data from the GfG study. We computed PGSs for various phenotypes (e.g., BMI, height, lifetime smoking, college education, etc.) for the 1,829 genotyped GfG participants who were age 50 and above and did not self-identify as Hispanic. GWAS meta-analyses for these phenotypes that produced the necessary PGS weights included hundreds of thousands of individuals (Wray et al., 2007). Our primary interest lies in estimating the relationships of the PGSs (our $Z$ variables of interest) with their corresponding measured phenotypes (our $Y$ variables of interest) and quantifying potential selection bias in these estimates.

Because applying the proposed indices of selection bias requires means, variances, and covariances for the covariates of interest $Z$ and the auxiliary variables $A$ for the non-selected cases, we used the Health and Retirement Study (HRS) as the source of auxiliary information for this target population. We computed these PGSs using *identical SNPs* for both GfG (our non-probability sample) and a benchmark probability sample (HRS) that collected the exact same genetic information and auxiliary variables $A$ (in this case, socio-demographics) measured in GfG. See the online supplementary materials for details regarding the common variables available in both the HRS and GfG, including the size of the HRS sample and the process used to determine SNPs that were measured in both studies. We estimated the means, variances, and covariances of the common $Z$ and $A$ variables in the target population (adults age 50 and above) using the HRS survey weights.

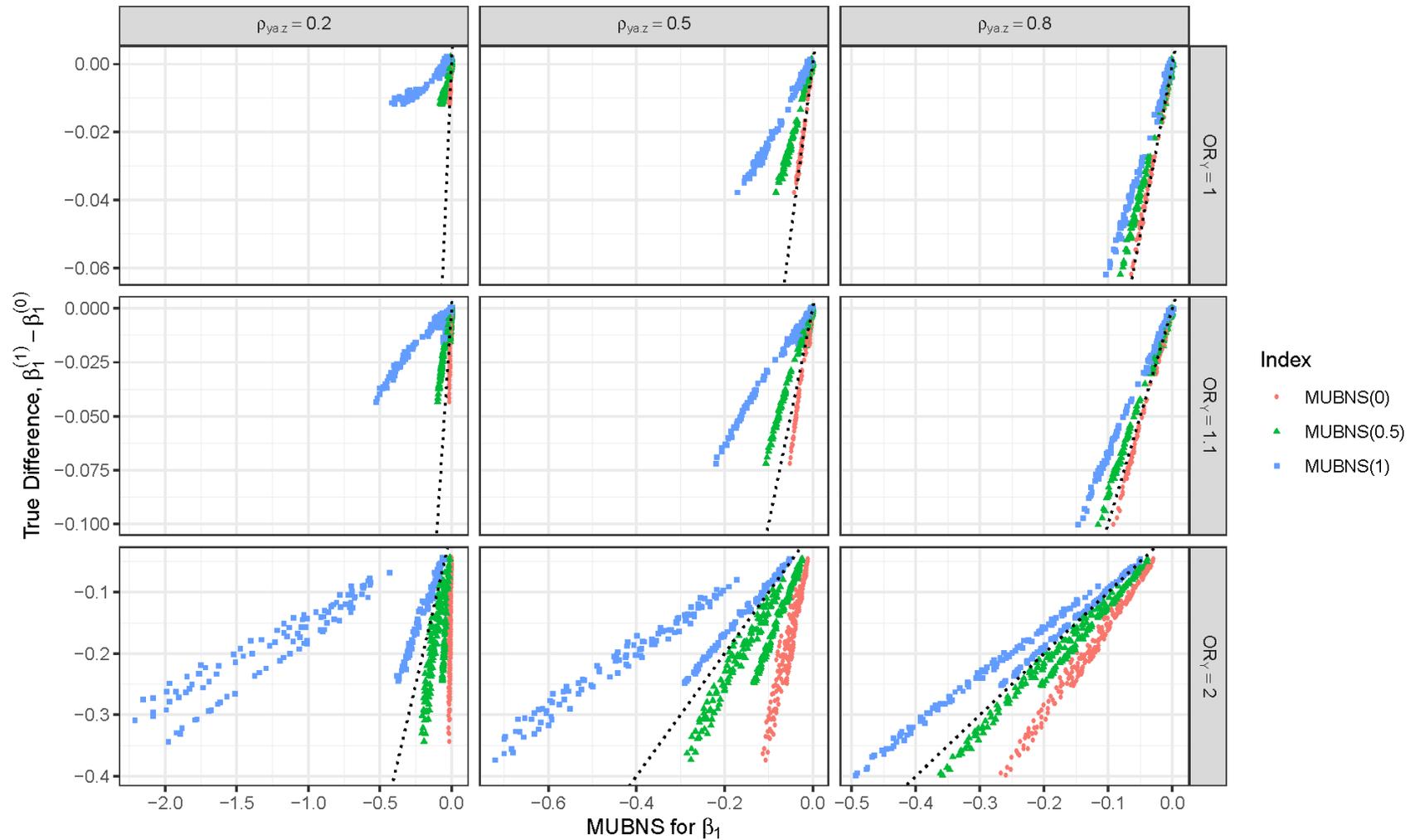

**Figure 1:** Scatter plots presenting associations between MUBNS and the true differences in coefficients between selected and non-selected units for the $Z_1$ coefficient. Results are median MUBNS values across 1,000 simulated datasets for each of the 1,944 combinations of data generation model and selection mechanism; panels are separated by the level of dependence on $Y$ in the selection model ($OR_Y$; rows) and the correlation between $Y$ and $A$ given $Z_1$ and $Z_2$ (columns). The dotted black line represents the $Y = X$ relationship.

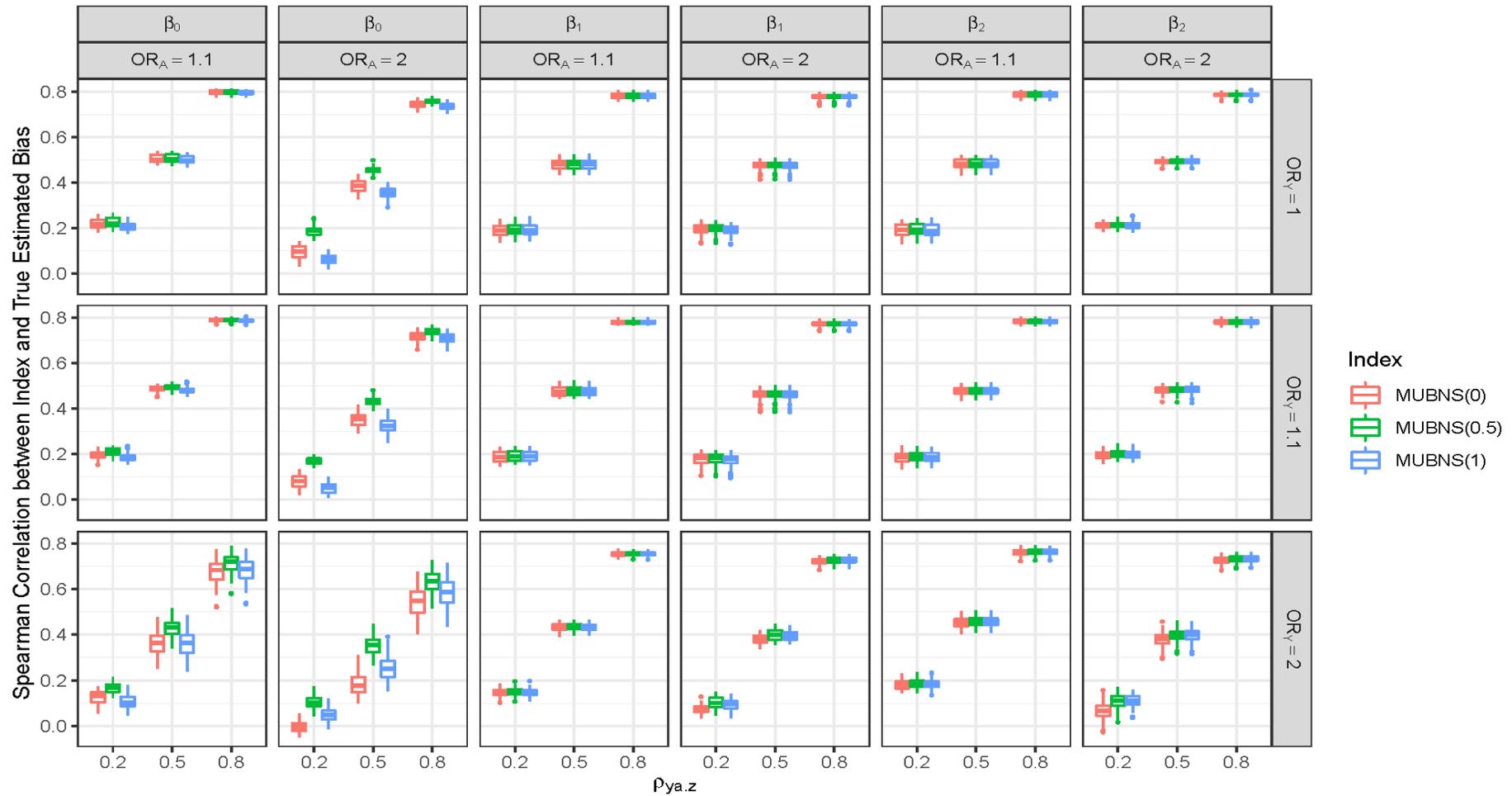

**Figure 2:** Side-by-side box plots presenting distributions of the Spearman correlations between MUBNS and the true difference in the coefficients between selected and non-selected units. We estimate each correlation from 1,000 replicate populations for each combination of data generation model and selection model. $OR_A$ = odds ratio for $A$ in the selection model; $OR_Y$ = odds ratio for $Y$ in the selection model.

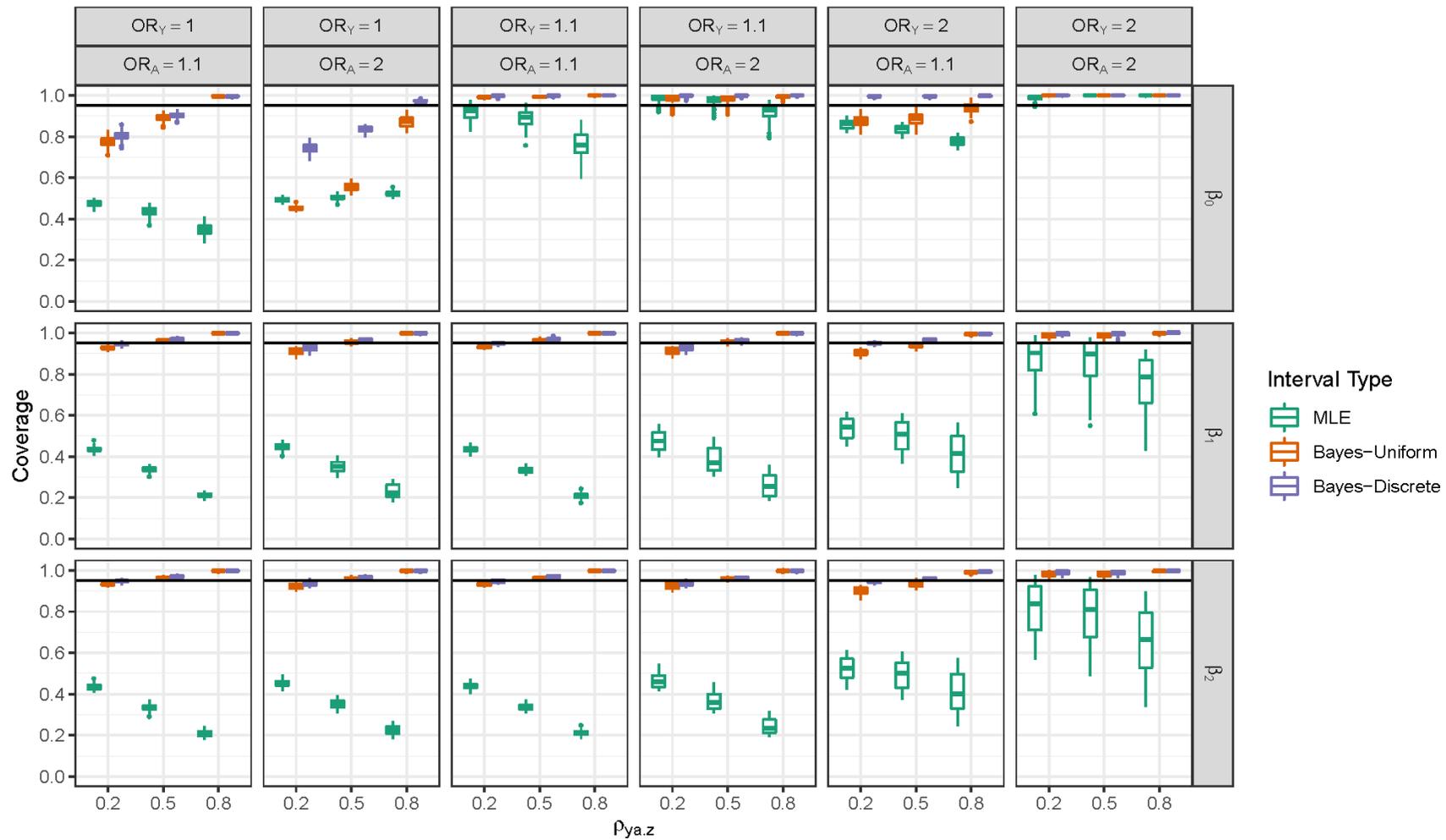

**Figure 3:** Side-by-side box plots presenting distributions of the empirical coverage rates for the alternative intervals. We estimate each coverage rate by computing the interval for each coefficient from 1,000 replicate populations for each combination of data generation model and selection model. $OR_A$ = odds ratio for A in the selection model; $OR_Y$ = odds ratio for Y in the selection model. The horizontal black line represents 0.95 coverage, for reference.

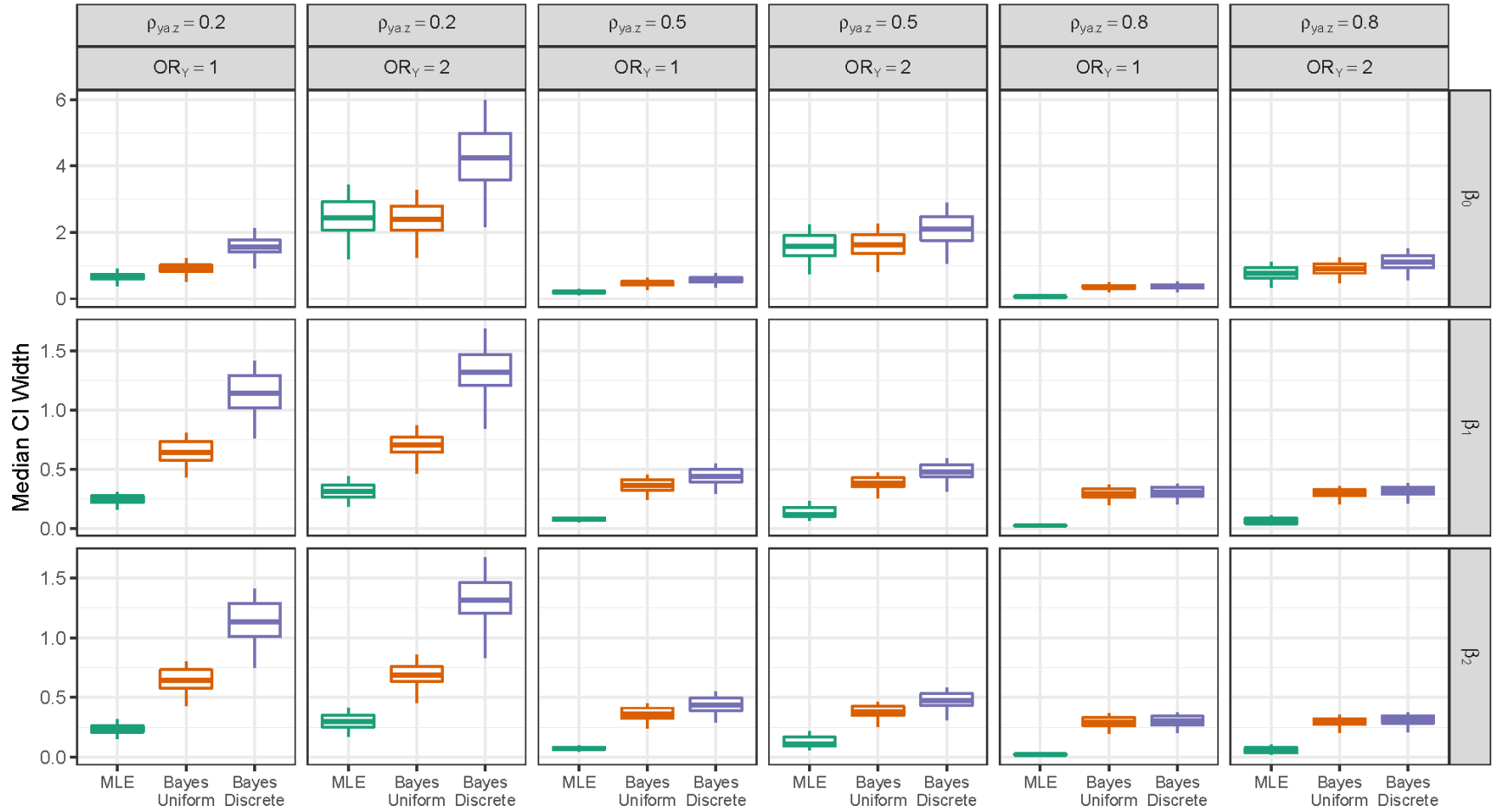

**Figure 4:** Side-by-side box plots presenting distributions of the empirical median widths for the alternative intervals across the different scenarios. We obtain the median width by computing the interval for each coefficient from 1,000 replicate populations for each combination of data generation model and selection model. $OR_A$ = odds ratio for $A$ in the selection model; $OR_Y$ = odds ratio for $Y$ in the selection model.

We then computed the MUBNS indices (given that the sampling fraction for GfG is unknown and likely quite small) for the coefficients of five linear regression models and three probit regression models fit to the GfG data, capturing potential bias in the estimated coefficients. For each model, the dependent variable $Y$ was a given continuous measure (height, BMI) or binary indicator (ever smoked more than 100 cigarettes, college degree [or greater], ever had diabetes). We analyzed the binary dependent variables using both linear and probit regression models to assess whether the indices proposed for probit regression models in Section 3.3 offered improved performance relative to the indices assuming a normal underlying pattern-mixture model. The corresponding mean-centered PGS for a given $Y$ variable was the $Z$ variable of primary interest in each model, and demographic measures (gender, education, birth cohort, age in years, race, and nativity) along with BMI and height (for all models aside from the BMI and height models) were the auxiliary variables $A$ used to compute $X$. As we mentioned above, the required aggregate summary statistics for $A$ (and therefore $X$) and $Z$ for the target population were estimated by performing survey-weighted analyses of the corresponding HRS data. The "true" values of the coefficients in each model arise from a fully design-based analysis of the HRS data, incorporating the complex sampling features (including weights) in estimation and variance estimation.

For the fully Bayesian approach to the analysis of the MUBNS indices for both the linear and probit regression models, we assumed a UNIFORM(0,1) prior for $\phi$ and non-informative Jeffreys' priors for the remaining parameters. We examined the correlation of the medians of the posterior draws of the MUBNS indices for each coefficient with their estimated biases, computed as the differences between the unweighted GfG coefficients and the survey-weighted estimates of the HRS coefficients. We also examined the ability of 95% credible intervals for the MUBNS indices to cover these estimated biases, and whether the intervals suggested a non-zero bias. Recall that the MUBNS index is based on the difference in a coefficient between the selected and non-selected cases. These analyses therefore assume a very small sampling fraction for the GfG cases, in which case the bias of the coefficient for the selected cases would be equal to the difference represented by the MUBNS index.

Table 1 presents the results of our analyses. Overall, we see that the estimates of bias in the intercepts and the PGS slopes based on the GfG data (when treating the HRS estimates as truth) are generally small, suggesting that selection bias in the GfG sample is not severe in the cases of these five models. We also note relatively small ($< 0.3$) conditional correlations of $X$ with $Y$ (when conditioning on the PGSs) for three of the five models, suggesting limited unique information in the additional auxiliary variables $A$ considered for these three models. The credible intervals for the MUBNS indices for the linear regression models cover the actual differences in the estimated coefficients between the GfG and the HRS in 7 out of 10 cases, and the credible intervals for the MUBNS indices for the probit regression models cover the actual differences in all six cases. This high coverage may be partly due to the wide intervals for the three models associated with the smallest conditional correlations (consistent with our simulation study).

**Table 1:** Estimates of coefficients in simple linear regression models for two continuous variables (height and BMI) and both simple linear regression models and probit regression models for three binary variables (ever smoke more than 100 cigarettes, college degree, and diabetes) as a function of the PGSs, from GfG (unweighted) and HRS (survey-weighted), in addition to posterior medians and 95% Bayesian credible intervals for the MUBNS index for each coefficient.

|  | GfG Coef. (SE): Linear / Probit | HRS Coef. (SE): Linear / Probit | Actual Est. Bias: Linear / Probit | Median of MUBNS posterior: Linear / Probit | 95% Credible interval for MUBNS: Linear / Probit | Cor(X,Y\|Z) |
|---|---|---|---|---|---|---|
| *Height* |  |  |  |  |  | 0.733 |
| Intercept | 66.07 (0.09) | 67.08 (0.09) | -1.01 | -2.87 | [-2.08, -3.90] |  |
| PGS slope | 0.82 (0.09) | 0.80 (0.15) | 0.02 | 0.40 | [-0.35, 1.24] |  |
| *Diabetes* |  |  |  |  |  | 0.324 |
| Intercept | 0.16 (0.01) / -0.99 (0.04) | 0.20 (0.01) / -0.87 (0.04) | -0.04 / -0.12 | 0.02 / -0.27 | [-0.03, 0.13] / [-1.29, 1.88] |  |
| PGS slope | 0.03 (0.01) / 0.13 (0.04) | 0.07 (0.01) / 0.26 (0.04) | -0.04 / -0.13 | -0.04 / 0.02 | [-0.14, -0.01] / [-0.28, 0.20] |  |
| *Ever Smoke* |  |  |  |  |  | 0.219 |
| Intercept | 0.62 (0.01) / 0.31 (0.03) | 0.58 (0.02) / 0.20 (0.05) | 0.04 / 0.11 | 0.00 / -0.44 | [-0.20, 0.17] / [-5.64, 2.91] |  |
| PGS slope | 0.05 (0.01) / 0.14 (0.03) | 0.01 (0.02) / 0.01 (0.05) | 0.05 / 0.13 | 0.09 / 0.18 | [0.01, 0.46] / [-0.05, 0.92] |  |
| *BMI* |  |  |  |  |  | 0.218 |
| Intercept | 29.65 (0.16) | 29.24 (0.18) | 0.41 | 1.84 | [0.37, 9.82] |  |
| PGS slope | 1.69 (0.16) | 1.79 (0.13) | -0.10 | 4.95 | [0.65, 27.49] |  |
| *College Degree* |  |  |  |  |  | 0.192 |
| Intercept | 0.51 (0.01) / 0.03 (0.03) | 0.37 (0.02) / -0.33 (0.04) | 0.14 / 0.36 | 0.28 / 1.82 | [0.06, 1.40] / [-0.49, 9.51] |  |
| PGS slope | -0.12 (0.01) / -0.31 (0.03) | 0.09 (0.01) / 0.23 (0.04) | -0.20 / -0.54 | -0.44 / -1.34 | [-2.93, -0.05] / [-5.73, 0.40] |  |

The Pearson correlations of the posterior medians for the MUBNS indices and the bias estimates in Table 1 were 0.56 for the five linear regression models and 0.86 for the three probit regression models, suggesting that these medians are useful indicators of potential bias, and could be used to order the coefficients in terms of their potential bias. Four GfG estimates present the strongest evidence of selection bias: the intercept in the model for height (corresponding to the expected height for the mean PGS), the PGS slope in the model for diabetes (more so for the linear regression model), and both the intercept and PGS coefficient in the models for the college degree indicator. The Bayesian credible intervals for the MUBNS indices provide correct evidence of a non-zero negative bias in the intercept and zero bias in the PGS slope from the linear regression model for the continuous height variable. This underscores the importance of informative auxiliary variables for the performance of the indices; note the wide intervals for the MUBNS indices that result from the low conditional correlation in the BMI model.

The credible interval for the MUBNS index for the PGS slope in the linear regression model fit to the binary diabetes indicator also correctly covers and provides evidence of the non-zero negative bias in the estimate of this slope. The credible interval for the MUBNS index for this same coefficient in the probit regression model includes zero, suggesting minimal selection bias for the GfG estimate. We note that this is actually consistent with the difference in the survey-weighted HRS coefficient and the GfG coefficient when fitting the probit regression models, where there is less statistical evidence of a difference in the coefficients (taking the standard errors into account) when compared to the linear regression model.

Finally, the MUBNS credible intervals for the coefficients in the linear regression model fit to the college degree indicator also provide correct evidence of non-zero positive and negative selection bias in the intercept and slope, respectively, despite the relatively small conditional correlation of $X$ with $Y$. The MUBNS intervals for the coefficients in the probit regression model fit to this indicator are also suggestive of positive and negative selection bias, although they slightly cover values of zero. In general, the intervals suggest that an adjustment of the inferences about these coefficients using the technique outlined earlier (Section 3.2, Remark 1) would repair this selection bias and provide a better sense of the positive relationship in this case.

Finally, we note that the MUBNS credible intervals are wider for the probit regression model in the case of this small conditional correlation, suggesting that these intervals may be more sensitive to these small correlations. We confirmed this with a replication of our simulation study using the indices for the probit regression model. We note that the MUBNS intervals for the intercepts in the probit regression models are expected to be wider than for the linear regression models fitted to the same binary indicators, given the wider range of the intercept for the underlying standard normal latent variable in the probit model specification.

We remind readers that we only needed sufficient statistics for the non-selected cases (estimated based on the HRS data) to compute the Bayesian credible intervals and that data from a large probability sample (e.g., HRS) could in general be employed to generate estimates of these quantities in other applications. Example code used for the calculations in this first application is available at https://github.com/bradytwest/IndicesOfNISB.

*5.2 Past-year Employment for Smartphone Users with Less Than High School Education in the National Survey of Family Growth*

As described in Section 2.2, we fit a linear regression model predicting the number of months worked in the past year as a function of gender (male / female) and age (15-18, 19-29, or 30-49), in the "non-probability sample" defined by smartphone users with less than high school education in the NSFG ($n$ = 2,977). We then computed the MUB indices and their intervals for the coefficients estimated from this subsample. We were able to compute the same coefficients for the "non-selected cases" in the remainder of the NSFG sample ($n$ = 16,823), enabling validation of the computed MUB indices. Our auxiliary variables in this application included race/ethnicity (non-Hispanic White, non-Hispanic Black, Hispanic, Other), marital status (married, divorced/widowed/separated), household income (<$19,999, $20,000-$59,999, $60,000+), region of the United States (Midwest, Northeast, South, and West), current employment status (working / not working), and presence of children under the age of 16 in the household (yes, no).

Table 2 presents the estimated coefficients in the model fitted to the non-probability sample, the same estimated coefficients in the model fitted to the full NSFG sample, the median of the MUB posterior distribution for each coefficient, and a 95% credible interval for MUB (where again the MUB captures potential bias in the estimated coefficients).

**Table 2:** Estimates of coefficients in simple linear regression models for the number of months worked in the past year as a function of gender and age, for both the non-probability sample defined by NSFG respondents who are smartphone users with less than high school education and the full NSFG sample (or "population"), in addition to posterior medians and 95% Bayesian credible intervals for the MUB index for each coefficient.

|  | Smartphone Users with Less Than HS Educ.: Coef. (SE) | Full NSFG "Population": Coef. (SE) | Estimated Bias | Median of MUB posterior distribution | 95% Credible interval for MUB |
| --- | --- | --- | --- | --- | --- |
| Intercept | 1.06 (0.13) | 2.09 (0.09) | -1.03 | -1.20 | [-1.87, -0.75] |
| Male | 1.34 (0.16) | 1.01 (0.07) | 0.33 | 0.44 | [0.16, 0.85] |
| Age 19-29 | 5.33 (0.20) | 5.64 (0.10) | -0.31 | -0.16 | [-0.63, 0.24] |
| Age 30-49 | 5.75 (0.18) | 6.43 (0.09) | -0.68 | -0.20 | [-0.67, 0.14] |

Compared to the population estimates based on the full NSFG sample, the estimates from the hypothetical non-probability sample suggest significantly lower mean past-year employment for younger females (the intercept term in each model). In addition, we see evidence of a larger estimated gap in the mean between males and females based on the non-probability sample, and smaller gaps between age groups 19-29 and 30-49 compared to those who are 15-18. The conditional correlation of the auxiliary proxy defined by *X* with number of months worked in the past year in this example was 0.692, which was nearly as high as that found for the height variable in the Genes for Good application. In this context, the posterior MUB medians had a high correlation with the actual differences in the coefficients between the selected and non-selected cases, and the 95% credible intervals for the MUB indices covered or nearly covered the actual differences in the coefficients between the non-probability sample and the full population without having extreme widths.

## 6. Discussion

We have addressed an important gap in the literature by developing model-based indices of selection bias for the coefficients in linear and probit regression models estimated from non-probability samples and evaluating the utility of these indices in different settings. Simulation studies and applications of the proposed measures to real data sets suggest that the indices are effective when informative auxiliary variables are available, especially the Bayesian version of the approach that takes into account uncertainty in the regression parameters for selected and non-selected cases. As Little et al. (2020) noted, quantifying non-ignorable selection bias for survey means and proportions may not be possible without access to informative auxiliary variables for the larger population. The same caveat applies to assessing selection bias in regression coefficients, with "informative" now meaning predictive of the outcome after conditioning on the covariates in the substantive model of interest. Without such auxiliary variables, no method is likely to be effective without strong structural assumptions about the regressions for the outcome or selection.

Collectively, our simulation study and our applications provide important recommendations for practice when applying these indices to assess potential selection bias in regression coefficients estimated from non-probability samples:
1. Identify good auxiliary predictors of the outcome variable in the model of interest, such that the correlation between a linear predictor of the outcome based on these auxiliary predictors and the outcome itself is moderate to high after conditioning on the primary predictor(s) of interest (e.g., the models for height and number of months worked in the past year in our applications);
2. If this conditional correlation is moderate to high, apply a fully Bayesian approach to form a credible interval for the measure of selection bias, namely MUB if the selection fraction is non-negligible and is known or can be estimated, and MUBNS otherwise; and
3. If this conditional correlation is low, the Bayesian credible intervals for the selection bias may become wide, reflecting the limited information available in the auxiliary variables used to form $X$.

We have provided code for computing the proposed MUBNS and MUB indices and forming both types of intervals at https://github.com/bradytwest/IndicesOfNISB.

Our method requires summary measures, either based on an external source of population information or a large probability sample, for auxiliary variables that are at least moderately predictive of the outcome $Y$, after adjusting for the covariates in the target regression model. We believe that such variables are necessary for any credible method for measuring selection bias. Public-use data files from large survey programs employing national probability samples, such as the HRS, provide good potential sources of this type of information.

This work has important implications for other studies in a variety of disciplines that are employing so-called big data, large volunteer samples, or convenience samples to make statements about relationships between variables in target populations, especially concerning genetics and genomics. In these situations, investigators do not have control over the selection mechanism that is generating the data. The indices proposed here can be used to assess the potential for selection bias in the estimated regression coefficients in such settings.

We employed a simple formulation of the polygenic score in our first application. Although commonly used in modern genetic research, PGSs have been criticized both from methodological and ethical angles. From a methodological perspective, missing heritability (differences in explained variability of disease occurrence between PGS and family studies) is a major limitation of the approach (Dudbridge, 2016; Wray et al., 2013). Even when including multiple genetic variables, the predictive power of PGSs is still very low and outperformed by simpler methods like family history (Dudbridge, 2016; Khoury, Janssens, and Ransohoff, 2013). Furthermore, SNPs included in the PGSs are often chosen using discovery thresholds based on *p*-values, which are known for their far-reaching limitations (Dudbridge, 2016; Maher, 2015; Wray et al., 2013), and final PGSs are obtained using somewhat arbitrary weighting of the SNPs (Maher, 2015). Another major critique of PGSs is the lack of representation of subjects with non-European ancestry (Lewis & Vassos, 2017; Torkamani, Wineinger & Topol, 2018). European ancestry subjects make up about 79% of all subjects in genetic studies, while this group represents 16% of the world's population. This disparity is expected to exacerbate existing health access disparities, given that methods are being developed for a population that already has better access to health services (Martin et al., 2019). The measures described in the present study will enable researchers to gauge potential selection biases in studies involving PGSs as predictors of other health outcomes.

Finally, future work in this area needs to extend the developments in this study to other types of generalized linear models (e.g., Poisson regression, generalized logit models, etc.). This would likely benefit applications where the dependent variables are not necessarily binary or normally distributed.